\documentclass[onecolumn,superscriptaddress,nofootinbib]{revtex4-2} 
\usepackage{geometry}
 \usepackage{dcolumn}    
\usepackage{pstricks}
\usepackage{color}
\usepackage{graphicx}
\usepackage{amsmath}
\usepackage{amssymb}
\usepackage[colorlinks=true,allcolors=darkred, pdfborder={0 0 0}]{hyperref}

\usepackage{booktabs}

\definecolor{darkred}{rgb}{0.5,0,0}
\definecolor{nicered}{rgb}{0.7,0.1,0.1}

\def\beqn#1{\begin{equation}\label{#1}}
\def\eeqn{\end{equation}}

\def\beqa#1{\begin{eqnarray}\label{#1}}
\def\eeqa{\end{eqnarray}}

\newcommand{\nc}{\newcommand}

\nc{\cO}{\mathcal{O}}
\nc{\cE}{\mathcal{E}}
\nc{\cD}{\mathcal{D}}
\nc{\cR}{\mathcal{R}}
\nc{\cP}{\mathcal{P}}
\nc{\cN}{\mathcal{N}}
\nc{\cL}{\mathcal{L}}

\newcommand\blfootnote[1]{%
  \begingroup
  \renewcommand\thefootnote{}\footnote{#1}%
  \addtocounter{footnote}{-1}%
  \endgroup
}
\begin{document}

\title{Uncovering hidden patterns in collider events with\\ Bayesian probabilistic models}

\author{Darius A. Faroughy}
\email{Proceedings for the $40$th International Conference on High Energy Physics - ICHEP 2020 Prague, Czechia}
\affiliation{Physik-Institut, University of Zürich\\ Winterthurerstrasse 190, CH-8057 Zürich, Switzerland}
%

\begin{abstract}
\vspace{0.3cm}

\begin{minipage}{13cm}

\noindent\rule{\textwidth}{0.3pt}

\vspace{0.1cm}

\noindent {\bf Abstract}\ \ Individual events at high-energy colliders like the LHC can be represented by a sequence of measurements, or `point patterns' in an observable space. Starting from this data representation, we build a simple Bayesian probabilistic model for event measurements useful for unsupervised event classification in beyond the standard model (BSM) studies. In order to arrive to this model we assume that the event measurements are exchangeable (and apply De Finetti's representation theorem), the data is discrete, and measurements are generated from multiple `latent' distributions, called {\it themes}. The resulting probabilistic model for collider events is a mixed-membership model known as {\it Latent Dirichlet Allocation} (LDA), a model extensively used in natural language processing applications. By training on point patterns in the primary Lund plane, we demonstrate that a two-theme LDA model can learn to distinguish in unlabelled dijet events the hidden new physics patterns produced by a BSM signature from a much larger QCD background. This note is based on refs.~\cite{Dillon:2019cqt,Dillon:2020quc}.
\noindent\rule{\textwidth}{0.3pt}

\end{minipage}

\vspace{0.3cm}

\end{abstract}

\maketitle

\section{Introduction}

A\blfootnote{\color{darkred}{Email: faroughy@physik.uzh.ch}} collider event $e$ can be represented by a sequence $\{o_1,o_2,\ldots, o_N\}$ of measurements, or observations $o_{i}$, taking values in some space $\cO$ spanned by a set of observables $\cO_1,\ldots,\cO_k$. For example, the $(p_T,\eta,\phi)$ particle coordinates at a hadron collider. Collider events can be thought as individual realizations of a {\it stochastic point process} in $\cO$. Each event can be represented by a distribution of points
\begin{align}\label{points}
e\,(o)\ =\ \sum_{i=1}^N\delta^{(k)}(o-o_i)\,,
\end{align}
\noindent in $\cO$, where the number $N$ of event measurements can be a random variables changing from event to event, or a deterministic quantity. For most collider events, the corresponding point patterns will not be uniformly distributed over $\cO$. For instance, at hadron colliders a substantial amount of the energy from the high-energy $pp$-collision is emitted in the form of collimated sprays of hadrons. These hadronic sprays, known as {\it jets}, lead to clustered point patterns in the space $\cO\!=\!(\eta,\phi)$. For high-level observables spanning $\cO$, the resulting point patterns for each event can be quite sparse, or give rise to irregularly shaped patterns when averaging over many events. For example, event point patterns in the Lund planes \cite{Dreyer:2018nbf} are both sparse and irregular in shape. Building a completely general probabilistic model for event measurements $\cP(e)=\cP(o_1,o_2,\ldots, o_N)$ for an arbitrary $\cO$ is therefore very challenging.\\ 

\section{A simple probabilistic model for collider events}
In this note we show that it is possible to write down a simple Bayesian probabilistic model for $\cP(e)$ that is capable of describing to a good approximation the generative process for event measurements. Moreover, following refs.~\cite{Dillon:2019cqt,Dillon:2020quc}, we demonstrate that the model can be used for unsupervised event classification. The probabilisitic model is based on three model-building assumptions: (i) Measurements in an event are exchangeable, (ii) the observable space $\cO$ is discretized, and (iii) event measurements are generated from multiple (latent) probability distributions over $\cO$.\\

\paragraph{{\bf Exchangeability.}} The first of the these assumptions requires that all event measurements are {\it exchangeable}, i.e. the order in which the measurements $o_i$  of an event are extracted is irrelevant. This implies permutation invariance: 
\begin{align}
\cP(o_1,\ldots, o_N) \ =\ \cP(o_{\pi(1)},\ldots, o_{\pi(N)})\,,
\end{align}

\noindent where $\pi$ is any element of the permutation group of $N$ indices. Exchangeability must not be confused with independent and identically distributed (iid). For iid measurements, the probability distribution would be completely factorizable and indeed exchangeable, but the converse wouldn't necessarily be true, not all exchangeable sequences are iid. Exchangeability actually implies a weaker notion of statistical independence called `conditional independence'. Both concepts are related through De Finetti's representation theorem:
\vspace{0.25cm}

\begin{center}
\fbox{
\begin{minipage}{11cm}
{\it {\bf De Finetti's representation  theorem:} A sequence of event measurements is exchangeable iff there exists a latent variable $\omega$ over some latent space $\Omega$, and a  distribution $\cP(\omega)$, such that 
\begin{align}\label{deFinetti}
\cP(o_1,\ldots, o_N)\ =\ \int_\Omega\, \mathrm{d}\omega\, \cP(\omega)\prod_{i=1}^N\,\cP(o_i|\omega)\,.
\end{align}}
\end{minipage}
\vspace{-0.3cm}
}
\end{center}

This result implies that if event measurements in $\cO$ are exchangeable, then these can be thought as being conditionally independent with respect to some marginalized hidden variable $\omega$. An event is generated by first sampling some random element $\omega$ from a latent space $\Omega$, then each measurement in the event is drawn from a distribution over $\cO$ conditioned on the drawn $\omega$. Looking closely at the integral representation in \eqref{deFinetti} one recognizes $\cP(\omega)$ as a prior and $\cP(o|\omega)$ as a likelihood, thus justifying the use of Bayesian probabilistic modelling for exchangeable data.\\

\paragraph{{\bf Measurement Discretization.}}  Permutation invariance leads to a very simple conditional structure for $\cP(o_1,\ldots, o_N)$, but De Finetti's theorem does not specify how to select the latent space $\Omega$, nor how to model the prior $\cP(\omega)$ or the conditional distribution $\cP(o|\omega)$ in \eqref{deFinetti}. For this, we need additional assumptions. One possibility, which makes parameter inference much simpler, is to choose the prior and likelihood to be conjugate distributions, for instance, these can belong to the exponential family. Our second model-building assumption is that the distribution $\cP(o|\omega)$ over $\cO$ is a discrete distribution. For this to make sense, we discretize the continuous observables spanning $\cO$ by binning this space so that the outcome of any event measurement is a discrete unit, or token, represented by the bin it populates. Notice that this `tokenization' of event measurements reduces the problem of finding a continuous distribution $\cP(o|\omega)$ over a multidimensional space $\cO$, to finding a discrete distribution over the finite set of non-negative integers labelling the bins in $\cO$. From all the discrete distributions in the exponential family, the most natural choice for $\cP(o|\omega)$ is the multinomial distribution (a multivariate generalization of the binomial distribution). This distribution is parametrized by a $M$-dimensional vector $\beta=(\beta_1,\cdots,\beta_M)$, satisfying
\begin{align}\label{simplex}
\sum^M_{m=1}\beta_m=1\ \ \ \text{and}\ \ \ 0\le\beta_m\le1,
\end{align}
where $M$ is the total number of bins that partition $\cO$ and the number $\beta_m$ represents the probability that a measurement $o_i$ populates the $m^\text{th}$ bin. In order to generate an individual (tokenized) event measurement $o_i$, we first draw $\omega$ from the prior, then, we randomly draw an index $m\in\{1,\ldots,M\}$ from the multinomial $\cP(o|\omega,\beta)$ conditioned on $\omega$. The resulting index points toward the bin  in $\cO$ that the measurement belongs to. The sampling of an event measurement from the multinomial can be pictured as rolling a dice with $M$ sides and bias $\beta$, which at this level is a free parameter of our probabilistic model. 

In order to `smooth' the multinomial parameter, we introduce a prior for $\beta$. The most natural prior is the {\it Dirichlet distribution}, a member of the exponential family that is conjugate to the multinomial distribution, defined as
\begin{align}
\cD(\beta|\eta)\ =\ \frac{\Gamma(\eta_1+\cdots+\eta_M)}{\Gamma(\eta_1)\cdots\Gamma(\eta_M)}\prod_{m=1}^M(\beta_m)^{\eta_m-1}\,.
\end{align}  
The Dirichlet $\cD(\cdot|\eta)$ is a family of distributions with {\it concentration parameter} $\eta=(\eta_1,\dots,\eta_M)$, $\eta_m>0$, and $\Gamma(x)$ denotes the Gamma function. The concentration parameter  controls the shape of the Dirichlet distribution over $\beta$ space. This space is an $(M-1)$-dimensional simplex. Notice that introducing this prior makes our model fully Bayesian, since we have replaced the task of fixing a large set of parameters (the probabilities $\beta$) of the multinomial with choosing a suitable Dirichlet distribution from which these parameters are sampled from. \\

\paragraph{{\bf Latent Dirichlet Allocation.} }We now need to specify the nature of the latent variable $\omega$ and the conditional dependence of the multinomial $\cP(o|\omega,\beta)$ with $\omega$. This brings us to our third  model-building assumption which is that the measurements $o_i$ in an event are assumed to arise from multiple multinomial distributions $\cP(o|t,\beta_t)$, labeled by a finite index $t\in\{1,\ldots,T\}$ and parametrized by $\beta_t=(\beta_{t1},\cdots,\beta_{tM})$. Each multinomial distribution represents an underlying event category, or {\it theme}, potentially describing features from multiple underlying physical processes or phenomena. ``Themes" is a terminology borrowed from the machine learning community, specifically from topic modelling and natural language processing. The latent variable is a $T$-dimensional vector $\omega=(\omega_1,\ldots,\omega_T)$ describing the relative proportion of every theme in the event. The likelihood in De Finetti's representation takes the form of a multinomial mixture model 
\begin{align}
\cP(o|\omega)\ =\ \sum_{t=1}^T\cP(t|\omega)\cP(o|t,\beta_t)\,.
\end{align}

\noindent The discrete distributions $\cP(t|\omega)$ are also multinomial distributions that are parametrized by the latent variable $\omega$. These represent the probability of selecting a particular theme $\cP(o|t,\beta_t)$ from which event measurements are extracted. The latent space $\Omega$ is a now a $(T-1)$-dimensional simplex, denoted by $\Omega_T$, spanned by the latent mixtures $\omega$ which now satisfy the convexity constraints as in \eqref{simplex}.\footnote{The simplex $\Omega_T$ must not be confused with the simplices for the multinomial theme parameters $\beta_t$.} This implies that the most natural choice for the prior $\cP(\omega)$ in \eqref{deFinetti} is the Dirichlet distribution over such simplex. 
With these model-building assumptions, we finally arrive to a fairly simple generative model for collider events over $\cO$:

\begin{align}
\cP(o_1,\ldots,o_N|\alpha,\eta)\ =\ \left(\prod_{t=1}^T\cD(\beta_t|\eta_t)\right)\,\int_{\Omega_T}\mathrm{d}\omega\,\cD(\omega|\alpha)\prod_{i=1}^N\left[\sum_{t=1}^T\cP(t|\omega)\,\cP(o_i|t,\beta_t)\right]
\end {align}

\noindent This model is known as Latent Dirichlet Allocation (LDA), and was first proposed as a topic model for texts\footnote{Other topic models have been previously used for collider studies in \cite{Metodiev:2018ftz} for quark/gluon jet discrimination.} \cite{Blei03latentdirichlet}. The model has two (multidimensional) model-building parameters governing the shapes of the Dirichlet distributions: the $T$-dimensional vector $\alpha=(\alpha_1,\ldots,\alpha_T)$ for the theme mixing proportions and a $T\times M$ matrix $\eta$ where the $M$-dimensional row $\eta_t$ controls the shape of the Dirichlet for the theme multinomials over $\cO$. The number of themes $T$ is also a model building parameter to be fixed before training these models with data. The simplest possible model is the two-theme LDA model. When $T=2$, the Dirichlet prior $\cD(\omega|\alpha_1,\alpha_2)$ becomes a beta distribution over the unit interval, and $\cP(t|\omega)$ is a binomial distribution over $t\in\{1,2\}$. After fixing the priors, the generative process for a single collider event goes as follows:

\begin{itemize}
\item (i) Draw a random mixing $\omega$ between zero and one from the beta prior.
\item (ii) Flip a coin with bias $\omega$.
\item (iii) If the coin lands on `heads'  select the first theme ($t=1$), otherwise select the second theme ($t=2$). 
\item (iv) Randomly sample one event measurement $o\in\cO$ from the selected theme multinomial by rolling an $M$-sided dice with bias $\beta_{t}$.
\item (v) Repeat steps (ii-iv) until all measurements $o_1,\ldots,o_N$ in the event have been generated.
\end{itemize}

LDA is a {\it mixed-membership model} because each measurements $o_i$ within an event can arise from multiple themes (e.g. a `head' or a `tail' theme when $T=2$), and each event within a sample exhibits these themes with different proportions. Mixed-membership models are not to be confused with classical mixture models. In the later, all measurements within an event are limited to come from a single theme (the mixture of theme is manifest at the event sample level, and not at the event level), while the former are more flexible probabilistic model that are capable of capturing common features between different underlying  physical processes.\\

\paragraph{{\bf Event classification with LDA.}}  After fixing the Dirichlet parameters $\alpha$, $\eta$ and the number of themes $T=2$, we can use LDA for unsupervised event classification. The posterior distribution $\cP(\omega,t,\beta|o_i,\alpha,\eta)$ is calculated using Bayes theorem. The idea is to learn from unlabelled collider data the theme multinomial parameters $\beta_{tm}$ and use them to cluster events into two categories. We use variational inference (VI) \cite{Blei03latentdirichlet} for the learning algorithm. During training, the algorithm learns the themes by identifying recurring measurement patterns, in particular, it identifies co-ocurrences between measurement bins throughout the event sample. Once the learning converges and the themes have been extracted, we build a likelihood-ratio defined by 

\begin{align}
\cL(o_1,\ldots, o_N|\alpha)\ =\ \prod_{i=1}^N\frac{\cP(o_{i}|1,\hat\beta_1(\alpha))}{\cP(o_{i}|2,\hat\beta_2(\alpha))}\,.
\end{align}

\noindent The $\hat \beta_t$ are statistical estimators for the $\beta_t$'s extracted from VI. The classifier is obtained by thresholding: for some suitable $c\in\mathbb{R}$, if $\cL(o_1,\ldots, o_N|\alpha)>c$ then the event belongs to theme $t=1$, else it belongs to theme $t=2$. This classifier is a function of the Dirichlet parameter $\alpha$, and is better thought as a continuous `landscape' of LDA classifiers. In principle there is no robust criteria for choosing one specific set of $\alpha$'s over another. Preliminary results given in ref.~\cite{Dillon:2020quc} suggest that a quantity known as {\it perplexity} can be used to precisely select the best $\alpha$.

\section{Latent Dirichlet allocation for jet substructure}

We now demonstrate how a two-theme LDA model can be used to uncover Beyond the SM (BSM) physics hiding in multi-jet events. First, we choose a set of jet observables for $\cO$. Observables that associate only one measurement to each event are not suitable for our method because this would produce for each event a single measurement\footnote{Jet substructure observables that marginalize over all particles in the event, like e.g. $N$-subjettiness \cite{Thaler:2010tr}, fall into this category and are therefore not useful for LDA.} in $\cO$. In order for LDA to learn from measurement co-ocurrence, we need observables that produce for every event a pattern of points in $\cO$. One possibility is to use observables extracted from the de-clustering history of jets. The jet de-clustering procedure generates a binary tree where each node corresponds to a splitting of a mother subjet into two subsequent daughter subjets $j_0\to j_1j_2$. During each splitting, a set of measurements $o$ is registered, generating a sequence of points in $\cO$ for the whole de-clustering tree. Assuming the de-clustering history to be exchangeable (i.e. ignoring the conditional dependence between measurements) is a good enough approximation for event classification purposes. For the splitting observables we choose  quantities that are sensitive to generic decay configurations of massive resonances, like the subjets invariant mass $m_0$, mass drop $m_1/m_0$, and Lund plane observables, $k_T$ and $\Delta$, defined in \cite{Dreyer:2018nbf}. We then build a multi-dimensional space $\cO$ spanned by different combination of these observables. Moreover, we also include a `jet label' indicating to which jet in the event the measurement was extracted from. \\

\begin{figure}[t!]
  \centering
  \includegraphics[width=0.5\textwidth]{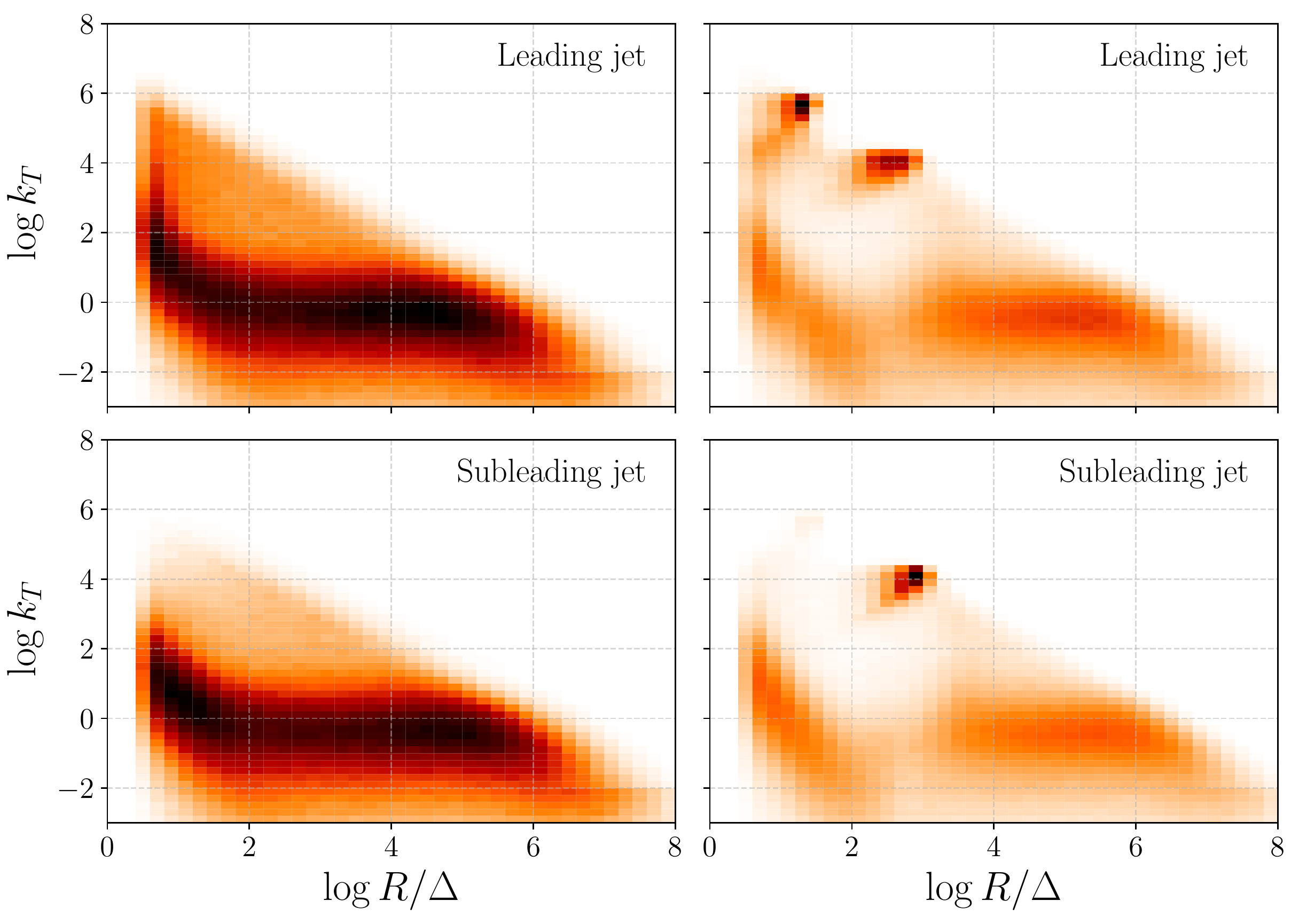}~  \includegraphics[width=0.5\textwidth]{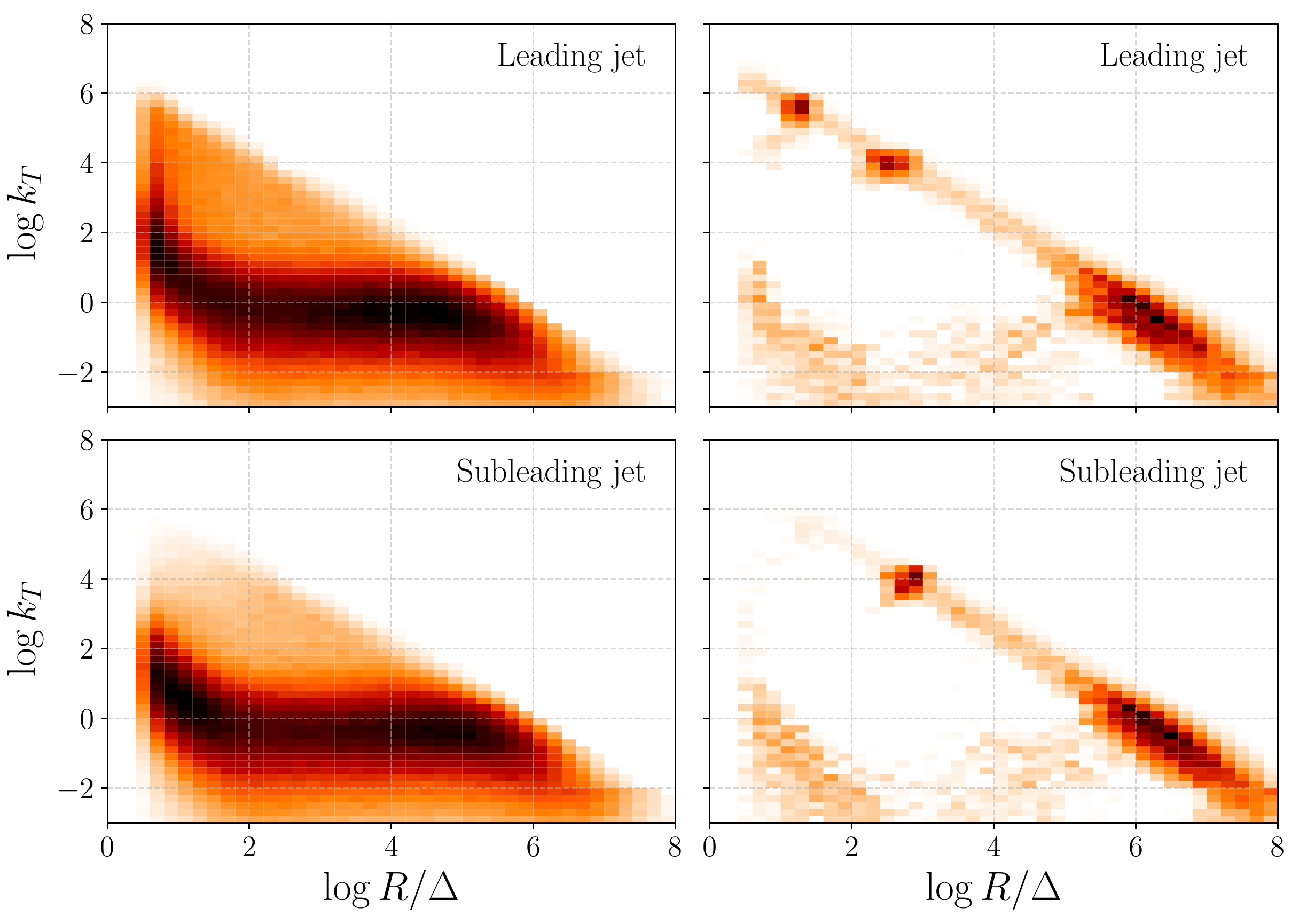}
  \caption{ \sl \small Truth-level primary Lund planes for QCD background ($1^\text{st}$ col.) and BSM signal ($2^\text{nd}$ col.). Results for the first theme ($3^\text{rd}$ col.) and second theme ($4^\text{th}$ col.), learned from a two-theme LDA model trained with 100k unlabelled events with $s/b=5\%$. See refs.~\cite{Dillon:2019cqt,Dillon:2020quc} for the full results.}
  \label{fig}
\end{figure}

In our studies, we used for the hidden BSM benchmark a $W^\prime\!-\!\phi$ model \cite{Dillon:2019cqt,Dillon:2020quc} with a boson mass $M_{W^\prime}=3$~TeV and scalar mass $M_\phi=0.4$~TeV. For the signal process we considered $pp\to W^\prime$ production followed by the decay chain $W^\prime \to W\phi\to WWW$, with $W$ bosons decaying hadronically. For the background we considered QCD dijet production. We generated 100k background and signal events and performed jet clustering using the $C/A$ algorithm with $R=1$. For the splitting observables we used the primary Lund plane $\cO$ spanned by $\{\log k_T, \log R/\Delta\}$, and also included labels $j=1,2,...$ indicating to which de-clustered jet in the event the measurement belongs (leading jet, subleading jet, etc..., ordered by invariant mass). The truth-level distributions for the primary Lund plane are given in figure~\ref{fig}, for the QCD background (first column) and signal (second column), for the leading jet (top row) and subleading jet (bottom row). The region near the hypothenuse of the Lund triangles describe the hard and collinear splittings. This region exhibits discriminating features between signal and background: for the signal we find two (one) dark clusters for the leading (subleading) jet, corresponding to the massive decay $\phi\!\to\! WW\!\to \!jjjj$ ($W\to jj$), while for the QCD background we expect a uniform pattern along the hypothenuse. We also can see non-perturbative features discriminating between background and signal along the $\log k_T\sim0$ axis. \\

We produced an unlabelled mixed sample of 100k events with $s/b=5\%$ and used it to train a two-theme LDA on the primary Lund plane with the {\tt Gensim} python package \cite{Rehurek10softwareframework}. For the Dirichlet prior $\cD(\omega|\alpha)$ controlling the theme mixings we fixed it to a very asymmetric shape $\alpha_0\approx0.9$ and $\alpha_2\approx0.1$. During training, this choice forces one multinomial theme ($t=1$) to approximate the mixed data distribution which we know (a priori) to be QCD-dominated because $s\ll b$. On the other hand, the other theme ($t=2$) is expected to learn non-QCD patterns in the Lund plane, with the hope that it picks up signal features. The outcome of the learned themes are shown in figure~\ref{fig}: the first theme (third column) matches very well with the QCD truth level distribution (first column), while the second theme (fourth column) contains the new physics signal features present in the truth level signal (second column). This result demonstrates that the two-theme LDA model can extract small BSM signals from a large background in a completely unsupervised manner. For more details see ref.~\cite{Dillon:2020quc}.

\section{Conclusions}
In conclusion, we have demonstrated that it is possible to build a simple generative probabilistic model for collider events. This model can be used for unsupervised event classification, e.g. for extracting BSM physics from jet substructure. The method presented here is based on a Bayesian probabilistic model called Latent Dirichlet Allocation. We arrived to this model starting from three main assumptions: (i) collider event measurements are to a good approximation `exchangeable', leading to De Finetti's integral representation for $\cP(o_1,o_2,\ldots)$, (ii) individual measurements are discrete (i.e. tokenized), and (iii) measurements arise from a multiplicity of latent (multinomial) distributions over $\cO$, called `themes'. We trained a two-theme LDA model on the primary Lund plane from a mixed dijet events sample with QCD background and BSM signal (a $W^\prime-\phi$ model) at $s/b=5\%$. Our results show that LDA can successfully discover small BSM signals in unlabelled data.

\bibliographystyle{JHEP}
\begingroup
    \setlength{\bibsep}{10pt}
    \linespread{1}\selectfont
    \bibliography{skeleton}

\providecommand{\href}[2]{#2}\begingroup\raggedright\begin{thebibliography}{1}

\bibitem{Dillon:2019cqt}
B.M.~Dillon, D.A.~Faroughy and J.F.~Kamenik, \emph{{Uncovering latent jet
  substructure}},
  \href{https://doi.org/10.1103/PhysRevD.100.056002}{\emph{Phys. Rev. D}
  {\bfseries 100} (2019) 056002}
  [\href{https://arxiv.org/abs/1904.04200}{{\ttfamily 1904.04200}}].

\bibitem{Dillon:2020quc}
B.~Dillon, D.~Faroughy, J.~Kamenik and M.~Szewc, \emph{{Learning the latent
  structure of collider events}},
  \href{https://doi.org/10.1007/JHEP10(2020)206}{\emph{JHEP} {\bfseries 10}
  (2020) 206} [\href{https://arxiv.org/abs/2005.12319}{{\ttfamily
  2005.12319}}].

\bibitem{Dreyer:2018nbf}
F.A.~Dreyer, G.P.~Salam and G.~Soyez, \emph{{The Lund Jet Plane}},
  \href{https://doi.org/10.1007/JHEP12(2018)064}{\emph{JHEP} {\bfseries 12}
  (2018) 064} [\href{https://arxiv.org/abs/1807.04758}{{\ttfamily
  1807.04758}}].

\bibitem{Metodiev:2018ftz}
E.M.~Metodiev and J.~Thaler, \emph{{Jet Topics: Disentangling Quarks and Gluons
  at Colliders}},
  \href{https://doi.org/10.1103/PhysRevLett.120.241602}{\emph{Phys. Rev. Lett.}
  {\bfseries 120} (2018) 241602}
  [\href{https://arxiv.org/abs/1802.00008}{{\ttfamily 1802.00008}}].

\bibitem{Blei03latentdirichlet}
D.M.~Blei, A.Y.~Ng, M.I.~Jordan and J.~Lafferty, \emph{Latent dirichlet
  allocation}, {\emph{Journal of Machine Learning Research} {\bfseries 3}
  (2003) 2003}.

\bibitem{Thaler:2010tr}
J.~Thaler and K.~Van~Tilburg, \emph{{Identifying Boosted Objects with
  N-subjettiness}}, \href{https://doi.org/10.1007/JHEP03(2011)015}{\emph{JHEP}
  {\bfseries 03} (2011) 015} [\href{https://arxiv.org/abs/1011.2268}{{\ttfamily
  1011.2268}}].

\bibitem{Rehurek10softwareframework}
R.~Rehurek and P.~Sojka, \emph{Software framework for topic modelling with
  large corpora},  in \emph{LREC 2010 WORKSHOP ON NEW CHALLENGES FOR NLP
  FRAMEWORKS}, pp.~45--50, 2010.

\end{thebibliography}\endgroup
\endgroup

\end{document}